\documentstyle[manuscript,aps]{revtex}
\begin{document}

\title{Finite Temperature Perturbation Theory and Large Gauge Invariance}

\author{Ashok Das}
\address{Department of Physics and Astronomy, University of Rochester,
Rochester, NY 14627}
\author{Gerald Dunne}
\address{Physics Department, University of Connecticut, Storrs, CT 06269}
\maketitle

\begin{abstract}
We examine finite temperature perturbation theory for Chern-Simons theories, in
the context of an analogue $0+1$-dimensional model. In particular, we show how
nonextensive terms arise in the perturbative finite temperature effective
action, using both the real-time and imaginary-time formalisms. We illustrate
how large gauge invariance is restored at all orders, despite being broken at
any given order in perturbation theory. We discuss which aspects generalize to
a perturbative analysis of finite temperature Chern-Simons terms in higher
dimensions.
\end{abstract}

\vskip .5in


\pagebreak
\section{Introduction}

Recently, some progress has been made in our understanding of the temperature
dependence of the induced Chern-Simons terms
\cite{dunne1,deser1,schaposnik1,aitchison1,gonzalez}. On the face of it, a
temperature
dependent induced Chern-Simons term would seem to violate gauge invariance
\cite{pisarski1,schaposnik2}, because a temperature dependent Chern-Simons
coefficient
cannot be chosen to take discrete values, as  invariance under large gauge
transformations would require
\cite{deser2}. However, in \cite{dunne1} a mechanism was demonstrated,
motivated by an exactly solvable $0+1$-dimensional Chern-Simons theory, whereby
the full effective action {\it does} satisfy large gauge invariance, in spite
of
the fact that it contains temperature dependent terms which violate large
gauge invariance at each order (of the field variable) in perturbation theory.

The essential new feature is that
at finite temperature, other parity violating terms (other than the
Chern-Simons term) can and do appear in the effective action; and if one takes
into
account all such terms to all orders (in the field variable) correctly, the
full effective action can maintain gauge
invariance even though it contains a Chern-Simons term with a temperature
dependent coefficient. In fact, it is clear that if there are higher order
terms present (which are not individually gauge invariant), one cannot ignore
them in discussing the question of invariance of the effective action under a
large gauge
transformation. Remarkably, this mechanism requires the existence of
nonextensive terms ({\it i.e.} terms that are not simply space-time integrals
of a density) in the finite temperature effective action, although only
extensive terms survive in the zero temperature limit. All these features have
been demonstrated explicitly and exactly in the $0+1$ Chern-Simons model
\cite{dunne1}. This idea has subsequently been analyzed in the framework of
zeta function regularization \cite{deser1}, and has been extended
to the Abelian $QED_{2+1}$ fermion determinant in special gauge backgrounds
that support gauge transformations with nontrivial winding numbers
\cite{deser1,schaposnik1,aitchison1,gonzalez}. Once again, the full effective
action is gauge invariant even though the induced Chern-Simons term has a
temperature dependent coefficient.

However, in our opinion, the issue of large gauge invariance, at finite
temperature, in a  $2+1$ dimensional Chern-Simons systems still poses many
interesting and
unsolved questions. The $QED_{2+1}$ fermion determinants studied in
\cite{deser1,schaposnik1,aitchison1,gonzalez} correspond to specially contrived
backgrounds that essentially factorize the effective action into the $0+1$
model of \cite{dunne1}
and a part corresponding to familiar 2 dimensional (Euclidean) fermion physics.
For a general gauge
background (and, more interestingly, for truly non-Abelian backgrounds) the
effective action cannot be computed exactly. Furthermore, Chern-Simons terms
may be induced not only in fermion systems, but also in purely gauge models
\cite{pisarski2,witten1}, and in gauge-Higgs models with spontaneous symmetry
breaking \cite{khlebnikov,chen} where, again, the exact evaluation of the
effective action may not be possible. Therefore, in such models, perturbation
theory (at finite
temperature) becomes a crucial and powerful tool. But, as stressed clearly in
\cite{deser1},
there is an inherent incompatibility between standard perturbation theory (at a
given order) and large gauge invariance, because the coupling constant cannot
be factored out of the large gauge transformation at finite temperature. The
{\it really} interesting question then, is to understand how to perform
reliable and consistent finite temperature perturbative calculations when large
gauge invariance is important. This is a very general question, and a deeper
understanding of this phenomenon should have important implications for finite
temperature QCD.

In this paper we make a modest first step in this direction by re-examining the
$0+1$ dimensional model of \cite{dunne1} using the various standard forms of
finite temperature perturbation theory \cite{das}. At first sight, it may seem
foolish to study perturbation theory for an exactly solvable model, but our
goal is to explore the intricacies of finite temperature perturbation theory in
the presence of large gauge invariance. We seek an understanding of how these
somewhat unfamiliar nonextensive terms (in the effective action) arise in
perturbation theory at finite temperature, and yet are all absent at zero
temperature. We also seek to identify precisely which features of this model
are special to $0+1$ dimensions, and which may be generalized to a perturbative
treatment of a $2+1$ dimensional model. In Section 2 we introduce the model and
review briefly the results of
\cite{dunne1,deser1,schaposnik1,aitchison1,gonzalez}. In
Sections 3 and 4 we use the real time formalism of finite temperature
perturbation theory to solve this system, in momentum space and then in
coordinate space. In Section 5 we apply finite temperature perturbation theory
in the imaginary time formalism. We conclude with some comments on the relative
merits of these approaches, and on which features may generalize to higher
dimensions.

\section{The Model}

Consider a $0+1$-dimensional field theory of $N_f$ flavors of fermions
$\psi_j$, $j=1\dots N_f$, minimally coupled to a $U(1)$ gauge field $A$. It is
not possible to write a Maxwell-like kinetic term for the gauge field in
$0+1$-dimensions, but we can write a Chern-Simons term - it is linear in $A$
\cite{csqm}. (Such $0+1$-dimensional Chern-Simons models have also been studied
recently in dimensional reductions of $2+1$-dimensional Chern-Simons theories
\cite{jackiw}). In ``Minkowski space'' (i.e. real time) the Lagrangian is
\begin{equation}
L=\sum_{j=1}^{N_f}\bar{\psi}_j  \left(i\partial_t-A-m\right)\psi_j
-\kappa A
\label{lag}
\end{equation}
This model supports gauge transformations with nontrivial winding number. Under
the $U(1)$ gauge transformation $\psi\to e^{-i \lambda}\psi$, $ A\to
A+\partial_t\lambda$, the Lagrangian changes by a total derivative and
the action changes by
\begin{equation}
\Delta S=-\kappa\int_{-\infty}^{+\infty} d t\,\partial_t \lambda=-2\pi \kappa N
\label{winding}
\end{equation}
where $N\equiv\frac{1}{2\pi}\int dt \partial_t \lambda$ is the
integer-valued `winding number' of the topologically nontrivial gauge
transformation. For example, the gauge transformation with
\begin{equation}
\lambda(t)=2N\, arctan\, t = -iN\, log\left({1+it\over 1-it}\right)
\label{large}
\end{equation}
has nonzero winding number $N$, and we see that $N$ must be an integer so that
the gauge transformation $\psi\to e^{-i\lambda}\psi$ preserves the
single-valuedness of the field $\psi(t)$.

Nevertheless, even though the classical action changes [as in (\ref{winding})]
under a large gauge transformation, the quantum path integral, which involves
$e^{i S}$, remains invariant provided the Chern-Simons coefficient $\kappa$ is
an integer. This is just the usual discreteness condition on the Chern-Simons
coefficient, familiar from three dimensional non-Abelian Chern-Simons
theories\cite{deser2}.

Induced Chern-Simons terms appear when we compute the fermion contribution to
the effective action for this theory:
\begin{equation}
\Gamma[A]=-i\log\left[{\det\left(i\partial_t- A-m\right)\over
\det\left(i\partial_t-m\right)}\right]^{N_f}
\label{effective}
\end{equation}
{}From \cite{dunne1}, we know the {\it exact} finite temperature effective
action for this theory to be (Note that under the Euclidean rotation $\int
d\tau\, A(\tau)\rightarrow -\int dt\, A(t)$.)
\begin{equation}
\Gamma[A]=-i N_f \log\left[\cos\left(\frac{1}{2}\int dt\, A\right)+i
\tanh\left(\frac{\beta m}{2}\right) \sin\left(\frac{1}{2}\int dt \,
A\right)\right]
\label{answer}
\end{equation}

Several comments are in order. First, notice that the effective action
$\Gamma[A]$ is not an extensive quantity ({\it i.e.} it is not an integral of a
density). Rather, it is a complicated function of the Chern-Simons action:
$\int
dt \,A$. Second, in the zero temperature limit, the effective action reduces to
\begin{equation}
\Gamma[A]_{T=0}=\frac{1}{2}{m\over |m|}N_f \int dt\, A(t)
\label{zero}
\end{equation}
which is the usual zero temperature induced Chern-Simons term. At nonzero
temperature the effective action is much more complicated. A formal
perturbative expansion of the exact result (\ref{answer}) in powers of the
gauge field yields
\begin{equation}
\Gamma[A]=\frac{N_f}{2}\left(\tanh\left(\frac{\beta m}{2}\right)
a+\frac{i}{4} {\rm sech}^2\left(\frac{\beta m}{2}\right) a^2+\frac{1}{12} \tanh
\left(\frac{\beta m}{2} \right) {\rm sech}^2\left(\frac{\beta m}{2}\right) a^3
+\dots\right)
\label{expansion}
\end{equation}
where we have defined
\begin{equation}
a\equiv \int dt\, A(t)
\label{cs}
\end{equation}
The first term in this perturbative expansion (\ref{expansion}) is the
Chern-Simons action, but with a temperature dependent coefficient, just as was
found in the perturbative computations in $2+1$ dimensional Chern-Simons
theories
\cite{babu,aitchison2}. If the computations stopped there, then we would arrive
at the apparent contradiction mentioned in the Introduction -- namely, the
``renormalized'' Chern-Simons coefficent
\begin{equation}
\kappa_R=\kappa-\frac{N_f}{2}tanh(\frac{\beta m}{2})
\label{ren}
\end{equation}
would be temperature dependent, and so could not take discrete values. Thus, it
would seem that the effective action cannot be invariant under large gauge
transformations. The flaw in this argument is clear. There are other terms in
the effective action besides the Chern-Simons term which cannot be ignored, and
these must all be taken
into account when considering the question of the large gauge invariance of the
effective action.
Indeed, the effective action (\ref{answer}) shifts by $(N_f\, N)\pi$,
independent of the temperature, under a large gauge transformation, for which
$a\to a+2\pi N$. This is just the familiar global
anomaly\cite{witten2,schwimmer}, which can be removed (for example) by taking
an even number of flavors, and is not directly related to the issue of the
temperature dependence of the Chern-Simons coefficient. The clearest way to
understand this global anomaly is through zeta function regularization of the
theory \cite{deser1}.

Finally, note that only the first term in the perturbative expansion
(\ref{expansion}) survives in the zero temperature limit. The higher order
terms are all nonextensive -- they are powers of the Chern-Simons action. The
corresponding Feynman diagrams vanish identically at zero temperature, and this
is usually understood by noting that they {\it must} vanish because there is no
gauge
invariant (even under infinitesimal gauge transformations) term involving more
than one factor of $A(t)$ that can be written down. This, however, assumes that
we only look for {\it extensive} terms; at nonzero temperature, this assumption
breaks down and correspondingly we shall see that our notion of perturbation
theory must be enlarged to incorporate nonextensive contributions to the
effective action. For example, let us
consider an action quadratic in the gauge fields which can have the general
form
\begin{equation}
\Gamma_{(2)} = \frac{1}{2}\int dt_{1}\,dt_{2}\,A(t_{1})F(t_{1}-t_{2})A(t_{2})
\label{quad}
\end{equation}
where we assume that $F(t_{1}-t_{2}) = F(t_{2}-t_{1})$. Under an infinitesimal
gauge transformation, this action will transform as
\begin{equation}
\delta \Gamma_{(2)} = -\int
dt_{1}\,dt_{2}\,\lambda(t_{1})\partial_{t_{1}}F(t_{1}-t_{2})A(t_{2})
\label{wardi}
\end{equation}
Clearly, the action will be invariant under an infinitesimal gauge
transformation if $F=0$. This corresponds to excluding the quadratic term
(\ref{quad}) from the effective action. But, as is clear from (\ref{wardi}),
the action can also be
invariant under infinitesimal gauge transformations if $F=constant$, which
would
make the quadratic action (\ref{quad}) nonextensive, and in fact proportional
to the square of the Chern-Simons action.
The origin of such nonextensive terms will be discussed in detail in the
following Sections
when we analyze this model using standard finite temperature perturbation
theory techniques.

To conclude this Section, we recall briefly the results of
\cite{deser1,schaposnik1,aitchison1,gonzalez} concerning the fermionic
determinant in
$QED_{2+1}$. In Euclidean space, using the imaginary time formalism, it has
been shown that for an Abelian gauge background of the form:
\begin{equation}
A_3={\rm constant}; \qquad \vec{A}(\vec{x}), \quad {\rm (indep.~of~time)}
\end{equation}
the parity odd part of the finite temperature effective action is
\begin{equation}
S_{odd}= i\, \Phi\,arctan\left[tanh(\frac{\beta m}{2})\,
tan(\frac{eA_3}{2})\right]
\label{deser}
\end{equation}
where $\Phi\equiv \frac{e}{2\pi}\int d^2x \, \epsilon_{ij} \partial_i A_j$ is
the time-independent magnetic flux of the background field. Notice that this
parity odd part of the effective action corresponds precisely to the real part
of the effective action (\ref{answer}), with the natural identifications
$A_3\to a$ and $\Phi\to N_f$.

\section{Real-Time Formalism: Momentum Space Calculation}

The perturbative computation of the fermionic contribution to the effective
action requires computing all
diagrams with one fermion loop and any number of external gauge fields. We
begin by considering the first few such diagrams for this
theory, in momentum
space, using the real-time formalism \cite{das}, before giving a systematic
method for evaluating them. The fermionic Feynman
propagator is (we assume from now on that $m > 0$, and since the propagator as
well as the vertices are diagonal in the flavor index, we do not write it
explicitly for simplicity)
\begin{equation}
S(p)=(p+m)\left({i\over p^2-m^2+i\epsilon}- 2\pi n_F(|p|)\delta(p^2-m^2)\right)
\label{propagator}
\end{equation}
where $n_F(|p|)$ is the Fermi statistical factor
\begin{equation}
n_F(|p|)={1\over e^{\beta|p|}+1}
\label{fermi}
\end{equation}
This propagator simplifies dramatically in $0+1$-dimensions, due to the trivial
one-dimensional nature of space-time. Using
$\delta(p^2-m^2)=\frac{1}{2m}[\delta(p-m)+\delta(p+m)]$, we find that
\begin{equation}
S(p)={i\over p-m+i\epsilon} -2\pi n_F(m)\delta(p-m)
\label{prop}
\end{equation}
Each fermion-gauge-fermion vertex contributes a factor of $-i$. Thus, the
contribution of the tadpole diagram to the linear term in the effective action
is (with the negative sign for the fermion loop):
\begin{eqnarray}
iI_{(1)}&=&-(-i)N_{f}\int{dp\over 2\pi} \left({i\over p-m+i\epsilon}- 2\pi
n_F(m)\delta(p-m)\right)\nonumber\\
&\equiv &i I_{(1)}^{(T=0)}+i I_{(1)}^{(\beta)}
\label{tad}
\end{eqnarray}
The zero temperature piece is
\begin{equation}
iI_{(1)}^{(T=0)}=-N_{f}\int{dp\over 2\pi} {1\over p-m+i\epsilon}=-N_{f}\int
{dp\over 2\pi}{p+m\over p^2-m^2+i\epsilon}=\frac{i}{2}N_{f}
\end{equation}
Note that to evaluate the first form of this integral directly, we must include
the contribution from the semicircle at infinity since the integrand does not
fall off fast enough. The temperature dependent tadpole contribution can be
evaluated trivially to give
\begin{equation}
iI_{(1)}^{(\beta)}=-2\pi i N_{f} n_F(m)\int {dp\over 2\pi} \delta(p-m)=-i\,
N_{f}n_F(m)
\end{equation}
Thus, the net tadpole diagram contribution is
\begin{equation}
iI_{(1)}=\frac{iN_{f}}{2}(1-2n_F(m))=\frac{iN_{f}}{2} tanh(\frac{\beta m}{2})
\label{tadpole}
\end{equation}
which gives a linear contribution to the effective action
\begin{equation}
\Gamma_{(1)}=-i\, A(k=0)\, iI_{(1)}=\frac{N_f}{2} tanh(\frac{\beta m}{2})\,
\int dt\, A(t)
\end{equation}
in agreement with the first term in the perturbative expansion
(\ref{expansion}) of the full effective action.

The two-point function also splits naturally into a zero
temperature piece and a temperature dependent piece (this is, in fact, a
general property of the real time formalism) and gives a contribution to the
quadratic term in the effective action of the form:
\begin{eqnarray}
iI_{(2)}(k)&=&(-){(-i)^2\over 2!}N_{f}\int{dp\over 2\pi}S(p)S(k+p)\equiv
iI_{(2)}^{(T=0)}(k)+iI_{(2)}^{(\beta)}(k)
\end{eqnarray}
The zero temperature piece is
\begin{eqnarray}
iI_{(2)}^{(T=0)}(k)&=&-\frac{N_{f}}{2}\int {dp\over 2\pi} \left({1\over
p-m+i\epsilon}\right)\, \left({1\over  (k+p)-m+i\epsilon}\right)\nonumber\\
&=&\frac{N_{f}}{4\pi}2\pi i\left[{1\over k}+{1\over -k}\right]\nonumber \\
&=&0
\end{eqnarray}
This is an explicit demonstration of the fact that the two-point function (and,
therefore, its contribution to the effective action)
vanishes identically at zero temperature, as required by (small) gauge
invariance. However, the finite temperature contribution is
\begin{eqnarray}
iI_{(2)}^{(\beta)}(k)&=&-\frac{N_{f}}{2}\int{dp\over 2\pi}\left[ 2\pi i n_F(m)
{\delta(p-m)\over (k+p)-m+i\epsilon}+2\pi i n_F(m){\delta(k+p-m)\over
p-m+i\epsilon}\right.\nonumber\\
&&\qquad \left.-4\pi^2 n_F^2(m) \delta(p-m) \delta(k+p-m) \right]\nonumber\\
&=&-\frac{i}{2}N_{f}n_F(m)\left[{1\over k+i\epsilon}+{1\over
-k+i\epsilon}\right]
+\pi N_{f} n_F^2(m)\delta(k)\nonumber\\
&=&-\pi N_{f}n_F(m)(1-n_F(m)) \, \delta(k)\nonumber\\
&=&-2\pi  \delta(k)\,\frac{N_{f}}{8}sech^2(\frac{\beta m}{2})
\label{ttwo}
\end{eqnarray}
Here we have used the identity $\delta(k)=\frac{1}{\pi}lim_{\epsilon\to 0}
{\epsilon\over k^2+\epsilon^2}$.
Since $I_{(2)}^{(T=0)}(k)$ vanishes, this result (\ref{ttwo}) gives the entire
two-point function. The resulting quadratic contribution to the effective
action is
\begin{equation}
\Gamma_{(2)}=-i\int{dk\over 2\pi}\, A(k)A(-k)\, i I_{(2)}(k) =i\frac{N_f}{8}
sech^2(\frac{\beta m}{2})\left(\int dt\, A(t)\right)^2
\end{equation}
in agreement with the perturbative expansion (\ref{expansion}).

There are two important things to observe from the structure of the two point
function: (i) its
dependence on the external momentum k is through a delta function $\delta(k)$;
and (ii) it
vanishes at zero temperature $(\beta\to\infty)$ because of the \lq\lq sech"
factor. The first
observation illustrates how nonextensive terms such as [cf. Eq. (\ref{quad})]
\begin{equation}
a^2 =\left(\int dt A(t)\right)^2=\int {dk\over 2\pi} A(k)A(-k) 2\pi \delta(k)
\end{equation}
arise in a perturbative approach to the finite temperature effective action,
while the latter explains why these are not seen in zero temperature
perturbation theory.

It is a straightforward matter to go ahead and evaluate diagrams with more than
two external gauge fields. However, motivated by the above result that the
two-point diagram is proportional to a delta function in the external momentum,
we appeal to the Ward identities for (small!) gauge invariance, which state
that the N-leg diagram (with $N\geq 2$), which is a function of $N-1$ external
momenta, satisfies the relations
\begin{equation}
k_j\, I_{(N)}(k_1,\dots,k_j,\dots k_{N-1})=0,\qquad {\rm for }\quad j=1,\dots
N-1
\label{ward}
\end{equation}
This is a generalization of (\ref{wardi}) to $N$-point functions in momentum
space
and here, there is no contraction of space-time indices since we are in a
one-dimensional
space-time! This implies that $I_{(N)}(k_1,\dots k_{N-1})$, for $N\geq 2$ must
be proportional to a product of delta functions in the $N-1$ external momenta:
\begin{equation}
iI_{(N)}(k_1,\dots,k_{N-1})= C_N(\beta m)\, \delta(k_1)\delta(k_2)\dots
\delta(k_{N-1})
\label{wi}
\end{equation}
where the coefficient $C_N$ is a function of $\beta m$ by dimensional
reasoning. But this immediately implies that the $N^{th}$ order contribution to
the effective action  in the perturbative expansion is proportional to the
$N^{th}$ power of the first order term, which is just the Chern-Simons action.
This is exactly the nonextensive structure that we observe in the perturbative
expansion (\ref{expansion}). Here we see that it is a direct consequence of the
Ward identities for small gauge invariance.

The result (\ref{wi}) suggests that the coefficients $C_N(\beta m)$
could be calculated from the $N$-leg diagram with zero external energies:
\begin{equation}
(-){(-i)^N\over N} N_{f} \int {dp\over 2\pi} \left[S(p)\right]^N
\label{singular}
\end{equation}
These are, however, extremely singular integrals because of the product of
delta functions with coincident arguments and have to be evaluated carefully.
We have done this, but it is rather complicated and we shall see in the
following Sections that it is much easier to evaluate these coefficients using
the coordinate space representation, or using the imaginary-time formalism.

\section{Real-Time Formalism: Coordinate Space Calculation}

In this Section we describe the coordinate space analysis of the perturbative
calculation in this model, using
the real-time formalism. The real-time propagator in the coordinate space can
be obtained simply from the Fourier transform of
the momentum space propagator (\ref{prop}):
\begin{equation}
S(t)\equiv \int{dp\over 2\pi}e^{-ipt} S(p)=
\left[\theta(t)-n_F(m)\right]e^{-imt}
\label{coordprop}
\end{equation}
Here $\theta(t)$ is the standard Heaviside step function.

Let us now consider the N-leg diagrams contributing to the effective action.
The contribution of the tadpole diagram is:
\begin{equation}
iI_{(1)}=-(-i)N_{f} S(0)=iN_{f}(\frac{1}{2}-n_F(m))=\frac{iN_{f}}{2}
tanh(\frac{\beta m}{2})
\label{coordone}
\end{equation}
The contribution of the two-point function to the quadratic action has the
form:
\begin{eqnarray}
iI_{(2)}(t_1,t_2)&=&-(-i)^2\frac{N_{f}}{2}S(t_1-t_2) S(t_2-t_1)\nonumber\\
&=&\frac{N_{f}}{2!}\left[\theta(t_1-t_2)-n_F(m)\right]
\left[\theta(t_2-t_1)-n_F(m)\right]\nonumber\\
&=&-\frac{N_{f}}{2}n_F(m)(1-n_F(m))\nonumber\\
&=&-\frac{N_{f}}{8} sech^2(\frac{\beta m}{2})
\label{coordtwo}
\end{eqnarray}
The three-point function gives a contribution of the form:
\begin{eqnarray}
iI_{(3)}(t_1,t_2,t_3)&=&-(-i)^3\frac{N_{f}}{3!}\left[S(t_1-t_2)S(t_2-t_3)
S(t_3-t_1) +
S(t_1-t_3)S(t_3-t_2)S(t_2-t_1)\right]\nonumber\\
&=& -\frac{iN_{f}}{6}\left[-n_F(m)+3n_F^2(m)-2n_F^3(m)\right]\nonumber\\
&=&\frac{iN_{f}}{6}n_F(m)(1-n_F(m))(1-2n_F(m))\nonumber\\
&=&\frac{iN_{f}}{24} tanh(\frac{\beta m}{2}) sech^2(\frac{\beta m}{2})
\label{coordthree}
\end{eqnarray}

We notice that the $N$-point functions are independent of the external
coordinates. This is the coordinate space analogue of the statement (\ref{wi})
that the momentum space $N$-point functions are proportional to products of
delta functions in the external momenta (or the generalization of (\ref{wardi})
to $N$-leg diagrams).
In coordinate space, it is easy to see
explicitly how this works. Clearly, the tadpole is independent of the external
coordinate. For the 2-point function (and its contribution to the quadratic
part of the action)
\begin{eqnarray}
{\partial\over \partial {t_1}}iI_{(2)}(t_1,t_2)&=& \frac{N_{f}}{2}
{\partial\over
\partial {t_1}}\left[(\theta(t_1-t_2)-n_F)(\theta(t_2-t_1)-n_F)\right]
\nonumber\\
&=&\frac{N_{f}}{2}\delta(t_1-t_2)\left[\theta(0)-n_F\right]-
\frac{N_{f}}{2}\delta(t_2-t_1)\left[ \theta(0)-n_F\right]\nonumber\\
&=&0
\end{eqnarray}
Similarly, for the 3-point function,
\begin{eqnarray}
{\partial\over \partial {t_1}}iI_{(3)}(t_1,t_2,t_3)&=&
-\frac{iN_{f}}{6}\delta(t_1-t_2)\left[S(t_2-t_3)S(t_3-t_1)-
S(t_1-t_3)S(t_3-t_2)\right]\nonumber\\&&\qquad
+\frac{iN_{f}}{6}\delta(t_1-t_3)\left[S(t_1-t_2)S(t_2-t_3)-
S(t_3-t_2)S(t_2-t_1)\right]
\nonumber\\
&=&0
\end{eqnarray}
In general, the derivative of the higher $N$-point functions with respect to
any one, say $t_1$, of the external coordinates vanishes because for any
diagram contributing to the $N$-point function with a given ordering of the
external coordinates, there is another diagram with $t_1$ interchanged with
another coordinate, say $t_2$. But ${\partial\over \partial {t_1}}
S(t_1-t_2)=-{\partial\over \partial {t_1}} S(t_2-t_1)$, and so these diagrams
cancel pairwise. We recognize that this is just a manifestation of the
coordinate space Ward Identities, which corresponds to a generalization of
(\ref{wardi}).

Furthermore, we notice from the results
(\ref{coordone},\ref{coordtwo},\ref{coordthree}) that $I_{(N)}$ is essentially
the derivative of $I_{(N-1)}$ with respect to $\beta m$, for $N\leq 3$. To
prove this in general, for any $N$, note that in any diagram contributing to
the effective
action, the product of the phase factors $e^{-imt}$ in a loop
simply cancel out. Therefore, we can effectively consider the `reduced'
propagator without this phase factor in our computations:
\begin{equation}
\tilde{S}(t_1-t_2)\equiv \theta(t_1-t_2)-n_F(m)
\label{red}
\end{equation}
Then it is clear that
\begin{equation}
{\partial\over \partial m}\tilde{S}(t_1-t_2)=-n_F^\prime(m)=-\beta\, n_F(1-n_F)
\end{equation}
where in the last step we have used the identity explicitly satisfied by the
Fermi factor in (\ref{fermi}). Since the Feynman amplitudes are independent of
the external time coordinates, we can choose any time ordering for these
quantities and we choose $t_1> t_2$. This gives,
\begin{equation}
{\partial\over \partial
{m}}\tilde{S}(t_1-t_2)=-\beta\tilde{S}(t_1-t_3)\tilde{S}(t_3-t_2), \qquad t_1>
t_2 >t_3
\end{equation}
where $t_3$ is arbitrary, but strictly less than $t_2$. Similarly,
\begin{equation}
{\partial\over \partial {m}}
\tilde{S}(t_2-t_1)=-\beta\tilde{S}(t_2-t_3)\tilde{S}(t_3-t_1), \qquad t_1> t_2
>t_3
\end{equation}
Thus, the effect of differentiating an $N$-point function with respect to $m$
is to introduce another coordinate (and, therefore, an external gauge field),
lower than all the others, in all possible
lines on the original diagrams. This is a generalization of the zero
temperature Ward identity and relates the $(N+1)$-point function to the
$N$-point
function through the following recursion relation:
\begin{equation}
{\partial\over \partial {m}} I_{(N)}=-i\beta (N+1) I_{(N+1)}
\label{rec}
\end{equation}
where we have suppressed the external coordinates since the $N$-point functions
are in fact independent of them.
But this means that {\it all} the coordinate space $N$-point diagrams are given
as derivatives of the tadpole diagram with respect to the mass $m$. Thus, the
full effective action is:
\begin{eqnarray}
\Gamma &=&-i\sum_{N=1}^\infty \left(\int dt A(t)\right)^N \,
(iI_{(N)})\nonumber\\
&=&-\frac{i\beta N_{f}}{2}\sum_{N=1}^{\infty} {\left(\frac{i}{\beta}\int dt
A(t)\right)^N \over N!} \, \left({\partial\over \partial m}\right)^{N-1}\,
tanh(\frac{\beta m}{2})
\label{resum}
\end{eqnarray}
Remarkably, this expansion may be resummed, to yield the full exact effective
action in (\ref{answer}).

\section{Imaginary Time: Momentum Space Calculation}

In this Section we present a perturbative analysis of the model using the
imaginary time formalism for finite temperature perturbation theory. The
coordinate space approach, in the imaginary time formalism, was already given
in \cite{dunne1}, so here we discuss the momentum space analysis.

Defining an imaginary time coordinate $\tau=i\, t$, the Lagrangian (\ref{lag})
becomes
\begin{equation}
L_{E}=\sum_{j=1}^{N_f}\psi^\dagger_j  \left(\partial_\tau-i A+m\right)\psi_j
-i\kappa A
\label{imlag}
\end{equation}
The imaginary time coordinate $\tau$ is restricted to the range $\tau\in
[0,\beta ]$, where $\beta$ is the inverse temperature. Fermi fields are
antiperiodic in $\tau$: $\psi(0)=-\psi(\beta)$; while gauge fields are
periodic: $A(0)=A(\beta)$. The propagators are the same as the zero temperature
propagators, but the antiperiodicity and periodicity conditions on the fields
imply that the corresponding energies take discrete values, being odd (even)
multiples of $2\pi T$ for fermions (bosons) \cite{das}.

The photon vertex, in this theory, contributes a factor of $i$ so that the
contribution of the `Euclidean' tadpole diagram to the effective action is
\begin{equation}
I_{(1)}=(-)(i)\,tr\left({1\over ip+m}\right)=-i\, tr\left({-ip+m\over
p^2+m^2}\right)
\end{equation}
There is no Dirac index except for the flavor index whose trace is trivial and
at zero temperature the energy trace is an integral, so
\begin{equation}
I_{(1)}= -iN_{f}\int_{-\infty}^\infty {dp\over 2\pi} \left({-ip+m\over
p^2+m^2}\right)
= -\frac{i}{2}\frac{m}{|m|}N_{f}
\label{imtad}
\end{equation}
Notice that the real part of the tadpole vanishes identically, while the
imaginary part is proportional to the sign of the mass $m$. At nonzero
temperature,
the trace is a sum over the discrete fermionic energies:
\begin{eqnarray}
I_{(1)}&=&(-iN_{f})T\sum_{n=-\infty}^\infty{-(2n+1)i\pi T+m\over
(2n+1)^2\pi^2T^2+m^2}
\nonumber\\
&=&-{iN_{f}m\over \pi^2 T} \sum_{n=-\infty}^\infty{1\over
(2n+1)^2+m^2\beta^2/\pi^2}
\nonumber\\
&=&-\frac{iN_{f}}{2}tanh(\frac{\beta m}{2})
\label{imtadpole}
\end{eqnarray}
The infinite sum, here, is just a standard  representation \cite{gradshteyn} of
the tanh function and this agrees exactly with the first term of eq. (14) in
\cite{dunne1}. In the zero temperature limit, this reduces smoothly to the zero
temperature result (\ref{imtad}) for the tadpole.

The `Euclidean' two-point diagram gives contribution to the quadratic action of
the form
\begin{equation}
I_{(2)}(k)=(-)\frac{(i)^2}{2!} \, tr\left({m^2-p(p+k)+i m(2p+k)\over
[m^2+p^2]\,
[m^2+(p+k)^2]}\right)
\label{euctwo}
\end{equation}
First consider this diagram at zero temperature. The imaginary part vanishes
identically, as can be seen by replacing $p$ by $-(k+p)$. The real part also
vanishes, as a result of a cancellation between the two terms:
\begin{equation}
\int_{-\infty}^\infty {dp\over 2\pi}\left({m^2-p(p+k)\over [m^2+p^2]\,
[m^2+(p+k)^2]}\right) ={|m|\over k^2+4m^2}-{|m|\over k^2+4m^2}=0
\label{cancellation}
\end{equation}
At nonzero temperature, the trace in (\ref{euctwo}) involves a sum over the
discrete fermionic
energies. Once again, the imaginary part vanishes after a shift of the energy
variable, but we note that this relies on the fact that the (bosonic) external
energy $k$ is an even multiple of $2\pi T$, while the (fermionic) loop energy
$p$ is an odd multiple of $2\pi T$. The real part requires considerably more
care, as we must distinguish between the case when the external energy vanishes
and when it is nonzero. When $k=0$, it is trivial to evaluate
\begin{eqnarray}
I_{(2)}(k=0)&=&\frac{N_{f}}{2} T\sum_{n=-\infty}^\infty{m^2-(2n+1)^2\pi^2
T^2\over
[m^2+(2n+1)^2\pi^2T^2]^2}\nonumber\\
&=& -\beta \frac{N_{f}}{8} sech^2(\frac{\beta m}{2})
\label{imtwo}
\end{eqnarray}
When $k\neq 0$, it is surprisingly tricky to evaluate this two-point diagram
(\ref{euctwo}) at finite temperature. A natural approach is to use Schwinger's
parametric representation of the integrand, in which case the diagram becomes a
parametric integral involving a Jacobi theta function. The advantage of this
approach is that the only difference between the zero temperature and nonzero
temperature result is the absence or presence of the theta function factor.
Represent the integrand as
\begin{eqnarray}
&&{m^2-p(p+k)\over [m^2+p^2]\, [m^2+(p+k)^2]}=[m^2-p(p+k)]\int_0^\infty
d\alpha_1\,\int_0^\infty d\alpha_2\,\, e^{-\alpha_1(m^2+p^2)}\,
e^{-\alpha_2 (m^2+(p+k)^2)}
\nonumber\\
&&\hskip 1cm =\int_0^\infty d\alpha_1\,\int_0^\infty d\alpha_2\,
\left[2m^2+\frac{1}{2}k^2+\frac{1}{2} {\partial\over \partial \alpha_1}+
\frac{1}{2} {\partial\over \partial \alpha_2}\right] e^{-\alpha_1(m^2+p^2)}\,
e^{-\alpha_2 (m^2+(p+k)^2)}
\end{eqnarray}
At zero temperature, the integration over the loop energy $p$ is simply a
Gaussian integral,
\begin{equation}
\int_{-\infty}^\infty {dp\over 2\pi} e^{-\alpha_1 p^2}\, e^{-\alpha_2(p+k)^2} =
{1\over \sqrt{4\pi(\alpha_1+\alpha_2)}}exp\left[-k^2\left({\alpha_1
\alpha_2\over \alpha_1+\alpha_2}\right) \right]
\label{gauss}
\end{equation}
After a change of variables
\begin{equation}
u=\alpha_1+\alpha_2; \qquad  v={\alpha_2\over \alpha_1+\alpha_2}
\end{equation}
for which the Jacobian equals $u$, we arrive at a simple parametric integral
representation for the two-point function:
\begin{eqnarray}
I_{(2)}^{(T=0)}(k)&=&\frac{N_{f}}{4\sqrt{\pi}} \int_0^1 dv\, \int_0^\infty udu
\left[ 2m^2+\frac{1}{2}k^2 +{\partial\over \partial u}+{1-2v\over
2u}{\partial\over \partial v}\right]\left(\frac{1}{\sqrt{u}}
e^{-u(m^2+v(1-v)k^2)}\right)\nonumber\\
&=&\frac{N_{f}}{4\sqrt{\pi}} \int_0^1 dv\, \int_0^\infty udu \left[
m^2+v(1-v)k^2-\frac{1}{2u}\right]\left(\frac{1}{\sqrt{u}}
e^{-u(m^2+v(1-v)k^2)}\right)\nonumber\\
&=&0
\label{imzero}
\end{eqnarray}
Thus, the two-point function vanishes at zero temperature, as in
(\ref{cancellation}).

At nonzero temperature, the energy trace is a sum, not an integral, so we
cannot perform a Gaussian integration as in (\ref{gauss}). Rather, the
summation
over the loop energy produces a Jacobi theta function. Write the loop energy as
$p=(2n+1)\pi T$ and the external gauge momentum as $k=2 l \pi T$. Then
\begin{equation}
\sum_{n=-\infty}^\infty e^{-\alpha_1 \pi^2T^2(2n+1)^2}\, e^{-\alpha_2 \pi^2 T^2
(2n+1+2l)^2} = e^{-4\pi^2 l^2 T^2\alpha_2} \, \Theta_2\left(4\pi i l
T^2\alpha_2|4\pi i T^2 (\alpha_1+\alpha_2)\right)
\end{equation}
where $\Theta_2$ is the second Jacobi theta function \cite{gradshteyn}:
\begin{equation}
\Theta_2(v|\tau)\equiv\sum_{n=-\infty}^\infty e^{i\pi \tau (n+1/2)^2}\, e^{i\pi
v(2n+1)}
\label{theta2}
\end{equation}
We now exploit the Poisson summation formula
\begin{equation}
\Theta_4\left(\frac{v}{\tau}|\frac{-1}{\tau}\right) =\sqrt{-i\tau}\, e^{i\pi
v^2/\tau} \, \Theta_2(v|\tau)
\label{poisson}
\end{equation}
where $\Theta_4$ is the fourth Jacobi theta function
\begin{equation}
\Theta_4(v|\tau)\equiv 1+ 2\sum_{n=1}^\infty (-1)^n e^{i\pi \tau n^2}\,
cos(2nv)
\label{theta4}
\end{equation}
Then, the finite temperature two-point function becomes
\begin{eqnarray}
I_{(2)}(k)&=&\frac{N_{f}}{4\sqrt{\pi}} \int_0^1 dv\, \int_0^\infty udu \left[
2m^2+\frac{1}{2}k^2 +{\partial\over \partial u}+{1-2v\over 2u}{\partial\over
\partial v}\right]\nonumber\\
&& \hskip 5cm \cdot\left(\frac{1}{\sqrt{u}} e^{-u(m^2+v(1-v)k^2)}
\Theta_4\left(\frac{k v}{2\pi T}|\frac{i}{4\pi T^2 u}\right)\right)
\label{im}
\end{eqnarray}
Remarkably, this expression only differs from the zero temperature parametric
expression (\ref{imzero}) by the presence of the theta function factor. When
$T\to 0$, this theta factor reduces to 1 and we get the zero temperature
expression. Furthermore, when the external
momentum $k$ vanishes, it is a straightforward exercise to show that we regain
the (nonzero) answer in (\ref{imtwo}). However, when $k\neq 0$, we can
integrate by parts
in $v$, and use the fact that the Jacobi theta functions satisfy a heat
equation
\begin{equation}
{\partial\over \partial\tau}\Theta(v|\tau)=-i\frac{\pi}{4}{\partial^2\over
\partial v^2} \Theta(v|\tau)
\label{heat}
\end{equation}
to convert the parametric integral (\ref{im}) into
\begin{eqnarray}
I_{(2)}(k\neq 0)&=&\frac{N_{f}}{4\sqrt{\pi}} \int_0^1 dv\, \int_0^\infty
\sqrt{u}\,
du\,  e^{-u(m^2+v(1-v)k^2)} \left[ m^2+v(1-v)k^2-\frac{1}{2u}-{\partial\over
\partial u}\right]\nonumber\\
&& \hskip 5cm \cdot \Theta_4\left(\frac{k v}{2\pi T}|\frac{i}{4\pi T^2 u}
\right)\nonumber\\
&=&0
\end{eqnarray}
Therefore, we have shown that the two-point function vanishes when the external
momentum $k$ is nonzero, but is nonzero when the external momentum $k$
vanishes. That is, the two-point function is proportional to a
Kronecker delta in the external momentum $k$:
\begin{equation}
I_{(2)}(k)=-\beta\,\delta_{k,0}\,\frac{N_{f}}{8}sech^2(\frac{\beta m}{2})
\label{kronecker}
\end{equation}
This is the imaginary time analogue of the real time result (\ref{ttwo}), and
it agrees exactly with the second term in Eq. (14) of \cite{dunne1}.
Similarly, using Schwinger's parametric representation, we can show that the
$N$-point function vanishes if any one of the external momenta is nonzero. This
is an explicit illustration of the Ward identity (\ref{wi}) in the
imaginary-time formalism.

When all the external momenta vanish it is very easy to evaluate the $N$-point
function using the imaginary time formalism:
\begin{eqnarray}
I_{(N)}(k_1=0,\dots,k_{N-1}=0)&=&(-) \frac{(i)^N}{N}N_{f}T
\sum_{n=-\infty}^\infty \left({1\over i(2n+1)\pi
T +m}\right)^N\nonumber\\
&=&\frac{N_{f}}{2}{(-i)^N\over N!}\left({\partial\over \partial m}\right)^{N-1}
tanh(\frac{\beta m}{2})
\label{imrec}
\end{eqnarray}
As before, this permits the effective action to be resummed:
\begin{eqnarray}
\Gamma_{E}&=&\sum_{N=1}^\infty T^{N-1}\sum_{k_1}\dots\sum_{k_{N-1}}A(k_1)\dots
A(k_{N-1}) A(-k_1-\dots-k_{N-1})\, I_{(N)}(k_1,\dots, k_{N-1}) \nonumber\\
&=&\sum_{N=1}^\infty T^{N-1}\, \left[A(k=0)\right]^N \, I_{(N)}(k_1=0,\dots,
k_{N-1}=0) \nonumber\\
&=&\beta \frac{N_f}{2}\sum_{N=1}^\infty {\left[-\frac{i}{\beta}\int_0^\beta
A(\tau)d\tau\right]^N\over N!} \left({\partial\over \partial m}\right)^{N-1}
tanh(\frac{\beta m}{2})
\end{eqnarray}
This agrees with the exact effective action derived in \cite{dunne1}, and is
the imaginary time version of the real time result (\ref{resum}).

\pagebreak

\section{Conclusions}

In conclusion, we have analyzed the fermion contribution to the effective
action in a
$0+1$-dimensional model with a Chern-Simons term, using various standard
techniques of finite temperature perturbation theory. Each formalism has its
own advantages and disadvantages. In the real-time formalism, the momentum
space calculation gives an immediate derivation, via the Ward identities
(\ref{wi}), of the existence of nonextensive terms in the effective action. But
the actual computation of the coefficients of these nonextensive terms is
rather messy in the momentum space approach. In particular, the diagrams with
zero external momenta are singular and must be treated with great care. Rather,
in the coordinate space approach (in the real time formalism) it is very easy
to evaluate these coefficients using the recursion relation (\ref{rec}) that
relates all higher order diagrams back to the tadpole diagram. It is also
rather straightforward to demonstrate the independence of the $N$-point
diagram on the external time coordinates, which is the coordinate space
statement of the Ward identities. Finally, the
momentum space calculation in the imaginary time formalism gives a trivial
evaluation of the $N$-point diagram when the external momenta are all zero.
However, it is surprisingly difficult to verify explicitly the Ward identities
by showing that the diagrams vanish if any
of the external momenta are nonzero.

Nevertheless, each of these different approaches works. The final result is (as
in \cite{dunne1}) that although each term in finite temperature perturbation
theory gives a
temperature dependent contribution that violates large gauge invariance, we can
resum all orders of perturbation theory to obtain the full effective action
(\ref{answer}), which respects large gauge invariance. We stress that gauge
invariance alone is not enough to determine the exact form of the effective
action.  Small gauge invariance implies (\ref{wi}) that the effective action
$\Gamma$ is a {\it function} of the Chern-Simons action $a$ ({\it i.e.} that
$\Gamma=\Gamma(a)$ is nonextensive). But to satisfy large gauge invariance all
we need is that the fermion determinant $e^{i\Gamma(a)}$ change by at most a
sign under a large gauge transformation: $a\to a+2\pi N$. This is satisfied by
a general expression
\begin{equation}
exp\left[ i\Gamma(a)/N_f\right] =\sum_{j=0}^\infty \left( d_j \,
cos(\frac{(2j+1)a}{2}) + f_j \, sin(\frac{(2j+1)a}{2})\right)
\end{equation}
The actual answer (\ref{answer}) gives as the only nonzero coefficients:
$d_0=1$ and $f_0=i\, tanh(\frac{\beta m}{2})$. This fact can only be deduced by
computation, not solely from gauge invariance requirements.

We now ask: which features of this $0+1$-dimensional model will extend to
the general $2+1$-dimensional case? In the $0+1$-dimensional case, the Ward
identities (\ref{ward}) have the simple consequence that nonextensive terms
will appear in the effective action (although the actual coefficients, zero or
nonzero, must be determined by a calculation). This argument does not
immediately generalize to $2+1$-dimensions, although it does apply to certain
special backgrounds such as the static ones used in
\cite{deser1,schaposnik1,aitchison1,gonzalez}; for these backgrounds the finite
temperature fermionic determinant can be calculated expicitly. However, the
nonextensive terms that appear are nonextensive in time only, and this is
probably a consequence of the fact that the backgrounds themselves are
static. Another dramatic
simplification of the $0+1$-dimensional model is that the propagator has a very
simple structure, both in momentum space (\ref{prop}) and in coordinate space
(\ref{coordprop}) primarily because there is  contribution only from the
positive energy terms. This is due to the fact that the on-shell condition is
very
simple for one-dimensional space-time, and also the fact that the Dirac spinor
structure is trivial. This simplicity is the key to deriving the recursive
relations (\ref{rec}) between the $N$-leg diagrams. These recursion relations
are essential for the resummation (\ref{resum}) of the perturbative expansion
to yield the full effective action (\ref{answer}), which is necessary to
demonstrate that large
gauge invariance is satisfied. Another simplifying feature of the
$0+1$-dimensional model is that there is no ambiguity in taking the zero
momentum limit of the $N$-leg diagrams (because there is just energy, no
spatial momentum). However, in higher dimensions the dependence of $N$-leg
diagrams on external momenta is nonanalytic at finite temperature
\cite{weldon}, and great care must be used in extracting
Chern-Simons-like terms via a zero external momentum limit. This problem simply
does not arise in the $0+1$-dimensional model. Finally, an important feature of
the $2+1$ induced Chern-Simons term at {\it zero} temperature is the
Coleman-Hill theorem \cite{coleman}, which essentially states that only
one-loop graphs contribute to the induced Chern-Simons term. But finite
temperature violates the assumptions used for the Coleman-Hill result, and it
is not clear what role this will play in a finite temperature perturbative
analysis of $2+1$-dimensional systems. However, this issue simply does not even
arise in the $0+1$-dimensional model, as the `photon' does not propagate; thus,
there are no higher loop diagrams to consider.

A direct perturbative analysis of $2+1$-dimensional models
will reveal whether these various simplifications of the $0+1$-dimensional
model are matters of convenience or if they are crucial to
restoring the large gauge invariance that finite temperature perturbation
theory breaks order by order \cite{dunne2}.

\vskip .5in
\noindent{\bf Acknowledgements}

This work was supported in part by US Department of Energy Grant No.
DE-FG-02-91ER40685 (AD) and Grant No. DE-FG-02-92ER40716 (GD).

\end{document}